# MCBs versus Fuses in Low-Voltage Applications: Critical Analysis Using the 3-AND Convergence Classifier

*Ishmael S. Msiza and Mehmood Haffejee*

Rietvlei Installations Department
Electrical Engineering Section, Rand Water
P.O. Box 1127 Johannesburg 2000, South Africa

imsiza@randwater.co.za, mhaffeje@randwater.co.za

**Abstract**

The protection of any electrical installation is one of the most important considerations in the field of electrical engineering. The electrical installation should be protected against hazards such as overload current and short circuit current conditions. The protection devices studied in this paper are the Miniature Circuit Breaker (MCB) and the High Rupturing Capacity (HRC) Fuse. Using a new tool known as the 3-AND Convergence Classifier, it was observed that the MCB is the most reliable in terms of offering protection against the damage caused by overload currents and the destruction caused by short circuit currents.

## 1   Introduction

Electrical installations have over the years demonstrated the need for protection from the hazardous effects of the over-current condition. This condition includes both overload currents and short circuit currents. Overload current flows for example, as a result of too many devices connected to the same energy source at the same time. Short circuit current flows when there is a zero resistance path to ground. It proved to be very useful during the formulation of Norton's theorem; hence it also became to be known as the Norton current [1]. The values of these currents are both larger than the rated current value of the device to be protected.

The primary objective of this paper is the analysis and comparison of MCBs versus fuses as protection devices in low voltage applications. The secondary objective of this paper is to demonstrate the feasibility of the new 3-AND Convergence Classifier.

This paper starts off by reflecting on some of the work that has been done in the field of evaluating the reliability of circuit protection devices. This is followed by a brief theoretical background of MCBs and fuses. The 3-AND Convergence Classifier is introduced, and then followed by its application. This paper concludes by discussing the results obtained from the analysis.

## 2   Related Works

The protection of electrical installations has been receiving a lot of attention due to the growing number of protection schemes and devices that are available for use. Some level of analysis has to go into the process of selecting the most optimum and reliable protection device for a particular application.

Work that has been done in this field includes the decision analysis worksheet that was developed by Sasol [2] to evaluate Moulded Case Circuit Breakers (MCCBs) versus Combined Fuse Switches (CFS) in some of their applications. The weakness of this worksheet is that it assigns a weight to each parameter using a scale of one to ten. This is prone to a lot of errors as the decision on the weight may vary from one person to the next. It is therefore better to have a weighting system that only has two states, similar to the one introduced in this paper.

Emanuelson *et al* [3] formulated a "from cradle to grave" life cycle assessment (LCA) of the miniature circuit breaker (MCB) versus the diazed fuse in household installations. This analysis was executed in order to determine the environmental impacts of these devices. A lot of assumptions went into this analysis and could have been improved by replacing the assumptions with real facts. This therefore implies that the need for the development of better and efficient analysis tools is sufficiently justified.

## 3   Theoretical Background

In terms of Rand Water's operations, low voltage (LV), alternating current (AC) applications are those applications with voltage levels less than and including 1000V. Rand Water's standardized voltages are 230V and 400V [4].

### 3.1   Miniature Circuit Breaker (MCB)

The MCB is an automatic, electrically operated switching device that was designed to automatically protect an electric circuit from overload currents and short circuit currents. It is a complicated construction made up of almost 100 individual parts [3]. It has the ability to respond within milliseconds when a fault has been detected. Westinghouse Electric introduced the World's first MCB and it initially had a porcelain base and cover mounted in a metal housing [5].

#### 3.1.1   Applications of MCBs

MCBs find wide application in residential, commercial and industrial operations. These applications include but are not limited to [6]:
- Power Supplies

- Programmable Logic Controller Input/Output (I/O) Points
- Lighting circuits
- Solenoids
- Relay/Contactor coils
- Appliances
- Control circuits
- Motor circuits

### 3.1.2 MCB Characteristics

The most essential feature of the MCB is the inverse-time tripping characteristic. This feature indicates the time required to trip the breaker in order to clear the circuit of any given level of overcurrent load. A typical inverse-time tripping characteristic is depicted in figure 1 below.

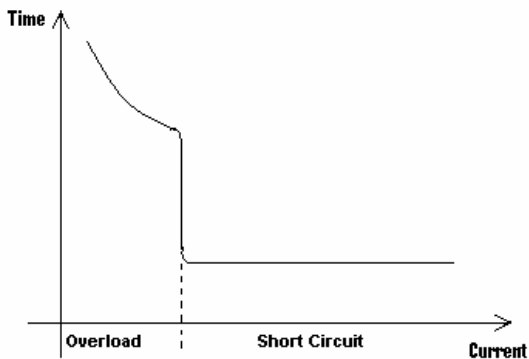

Figure 1: Inverse-time tripping feature of the MCB

### 3.1.3 MCB Operation

The operation of the MCB in order to ensure protection against overload currents and short circuit currents is summarized in figure 2 below.

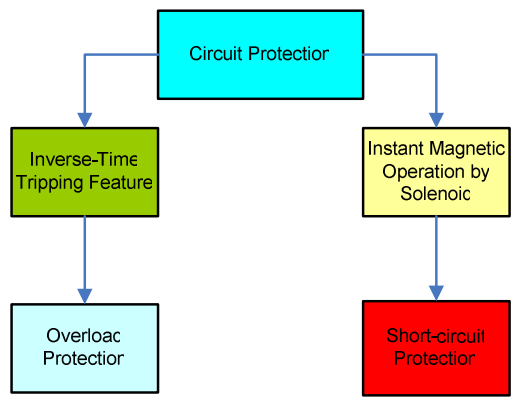

Figure 2: The operation of the MCB

### 3.1.4 MCB Advantages over fuses

The advantages of the MCB can be summarized as follows:
- Closed overload protection compared to HRC fuses
- Stable tripping characteristics
- Common tripping of all the phases of a motor
- Instant re-closing of the circuit after a fault has been cleared
- Safety disconnect features for circuit isolation
- Terminal insulation for operator safety
- Ampere ratings that can be fixed and modified compared to the possibility of introducing overrated fuses
- It is reusable, hence very little maintenance and replacement costs
- Lower power losses
- Simplicity of mounting and wiring
- Lower space requirements
- Provision of accessories e.g. auxiliary switch
- Stable arc interruption
- Discrimination can be achieved either based on current or based on time

### 3.1.5 MCB Disadvantages

The disadvantages of the MCB can be summarized as follows:
- More expensive than the fuse
- Difficult to identify where the fault occurred
- Fault can be cleared in any time up to 10 cycles of the current waveform
- Large amount of energy "let through" (10 times that released by the fuse)

## 3.2 High Rupturing Capacity (HRC) Fuse

The word fuse is a short form of "fusible link" and it is also protection device capable of protecting a circuit from overload currents and short circuit currents. In this paper, the fuse is viewed as part of a switch and hence can be referred to as a fuse switch. Fuses are rated in terms of many aspects. These include voltage, current and the type of application. A high rupturing capacity (HRC) fuse is a fuse that has a high breaking capacity (higher kA Rating). The minimum fault value for an HRC fuse is 80kA. A fuse should be selected with a rating just above the normal operating current of the device to be protected. A general approach is that it should operate at 1.2 times the rated current. A typical fuse is made of silver-coated copper strips and granular quartz [7].

### 3.2.1 Fuse Applications

Fuses find application in systems where the load does not vary much above the normal value (overload protection). They also find application in systems where the loads vary considerably (short-circuit protection). These applications include but are not limited to [8]:
- Transformer circuits
- Capacitor banks
- Motor circuits
- Fluorescent lighting circuits
- Control circuits

### 3.2.2 Fuse Characteristics

The inverse time-current characteristic shows the time required melting the fuse and the time required to clear the circuit for any given level of over current load. A simplified, but typical fuse time-current feature is depicted in figure 3 below.

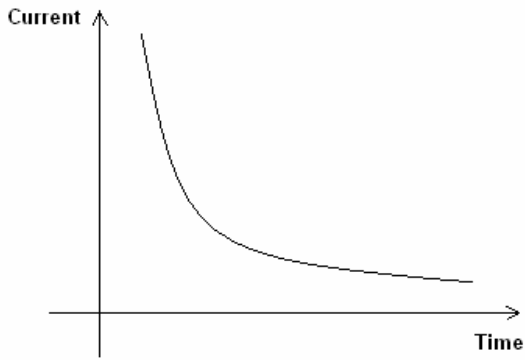

Figure 3: Typical time-current feature of a fuse

This curve is very important when determining an application for a fuse as it allows the correct ratings to be chosen.

### 3.2.3 Fuse Operation

When an over current condition occurs in the circuit, the silver-coated metal strip melts. It subsequently melts the surrounding quartz and this combination forms an insulating material called fulgarite [7]. Like any perfect insulator, fulgarite has an infinite resistance and hence it creates an open circuit. This operation is summarized in figure 4 below.

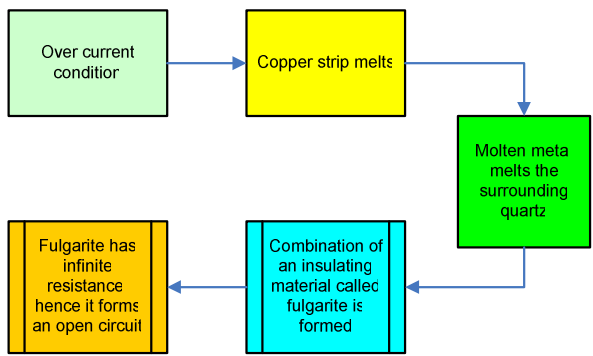

Figure 4: The operation of the fuse

### 3.2.4 Fuse Advantages over MCBs

The advantages of the fuse can be summarized as follows:
- Cheaper when compared to MCBs
- It is easy to identify where the fault is due to the open air gap
- It can cut-off fault current long before it reaches its first peak
- Hence, very little energy "let through" ($I^2t$)
- Perfect discrimination easily achievable due to the low cut-off value

### 3.2.5 Fuse Disadvantages

The disadvantages of the fuse can be summarized as follows:
- The abrupt introduction of high resistance in the circuit by a badly designed and assembled fuse can create unwanted effects while clearing the fault
- Although this is very rare, fuses are likely to produce high peak voltage which is much higher than the system voltage and can puncture the insulation of the rest of the circuit
- A lot of maintenance and replacement costs. Maintenance in the form of continuously monitoring the state of the fuse; and replacement after each and every fault
- The cut-off current increases with the fuse rating
- Fuse of incorrect ratings can easily be installed in the fuse holders
- In a three phase power circuit, if one fuse blows, all the fuses must be replaced at the same time

## 4 The 3-AND Convergence Classifier

### 4.1 Detailed Description

The 3-AND Convergence Classifier is an innovative tool that can be used to classify an object either as good or bad, when it is compared with another object for a particular application. This classifier has not been used before; hence it is first introduced and implemented in this paper.

This classifier has a discrete nature because it only has two states to represent an outcome. These states are the 0-state and the 1-state, where the 0-state indicates that an object has been classified as bad when compared with the second one. Consequently, the 1-state indicates that an object has been classified as good when measured against its counterpart. In this paper, the two objects are obviously the MCB and the HRC Fuse. This classifier consists of three layers; the analysis layer, the convergence layer and the output layer, as shown in figure 5 below.

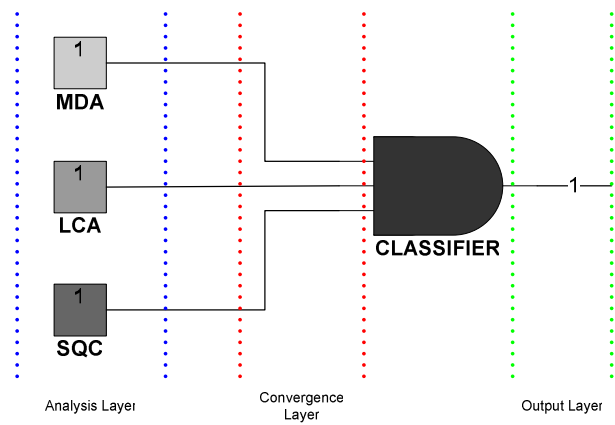

Figure 5: The 3-AND Convergence Classifier

The analysis layer encapsulates a total of three analysis paradigms. These analysis paradigms may vary from one application to the next. In this paper, the analysis paradigms considered are the Merit and De-merit Analysis (MDA), Life Cycle Assessment (LCA) and Speed, Quality and Cost (SQC) evaluation.

### 4.2 Algorithmic Description

**Algorithm 1: The 3-AND Convergence Classifier**

**do**
  →Compare the two objects in each and every analysis paradigm
**enddo**

**if** an object satisfies a paradigm
    **then** it is allocated the 1-state for that particular paradigm
    **else** it is allocated the 0-state
**endif**

**do**
  →Converge all the outcomes of the individual paradigms
**enddo**

**if** all the outcomes of the individual paradigms are of 1-state for the same object
    **then** that object is classified as good
    **elseif** all or one of the outcomes of the individual paradigms is of 0-state for the same object
        **then** that object is classified as bad
    **endelseif**
**endif**

### 4.3 Classifier input and output

The 3-AND Classifier has a total of three inputs and a single output. The inputs are the results from the individual analysis paradigms. They are converged into a model of an AND gate and classified as either good or bad. The input-output functional mapping is given by (1).

$$y = x1 \, AND \, x2 \, AND \, x3 \qquad (1)$$

Where $y$ is the output, $x1$ is the result of the first analysis paradigm, $x2$ is the result from the second analysis paradigm and $x3$ from the third.

Since the classifier is modeled as an AND gate, it is possible to use the concept of digital electronics [9] to represent the input-output functional mapping. In other to represent this functional mapping, a truth table is as depicted below.

Table 1: Truth table of the 3-AND Convergence Classifier

| $x1$ | $x2$ | $x3$ | $y$ |
|---|---|---|---|
| 0 | 0 | 0 | 0 |
| 0 | 0 | 1 | 0 |
| 0 | 1 | 0 | 0 |
| 0 | 1 | 1 | 0 |
| 1 | 0 | 0 | 0 |
| 1 | 0 | 1 | 0 |
| 1 | 1 | 0 | 0 |
| 1 | 1 | 1 | 1 |

It is evident from the truth table that the an object can only be classified as good if and only if it passes the all the individual analysis paradigms.

### 4.4 Properties of the Classifier

The properties taken into consideration are similar to those considered in the study of signal and system analysis [10]. These properties are memory, invertibility, causality, stability and linearity.

This classifier is memoryless because an output at $n_0$, $y[n_0]$ does not depend on input values other than $x[n_0]$. It is also non-invertible because distinct inputs do not produce distinct outputs. This can be observed from the truth table. The output column has more than one zero-state from many input combinations.

This classifier is not causal because it is not a physical system; it is just an analysis tool. In terms of bounded input bounded output (BIBO) stability, this classifier is stable because the output remains bounded (0 or 1) for any bounded input (0 or 1). It is not linear because it fails both the tests of additivity and homogeneity.

### 4.5 Merits of the Classifier

The only limitation of the 3-AND Convergence Classifier is that it can only be employed to compare two objects. The advantages of using this classifier can be summarized as follows:
- It is a stable system
- It is a discrete system
- Sampling of inputs is unnecessary as they only have two possible states
- It is unaffected by time-invariance
- It is not complex (only $2^3$ possible input combinations)

## 5 Analysis Using the Classifier

The MCB and the HRC Fuse are now analyzed using the 3-AND Convergence Classifier.

### 5.1 Merit and De-merit Analysis (MDA)

This analysis compares the advantages and disadvantages of both objects. From section 3.1.4 and 3.2.4 it appears that the MCB has a total of 14 advantages over the fuse while the fuse only has 5 advantages over the MCB. From section 3.1.5 and 3.2.5 it appears that the fuse has a total of 6 disadvantages while the MCB only has a total of 4.

**Fuse MDA Score:** 0
**MCB MDA Score:** 1

### 5.2 Life Cycle Assessment (LCA)

#### 5.2.1 Lifespan
The fuse generally has a shorter lifespan than the MCB [3]. This implies that the MCB lives longer than the fuse.

#### 5.2.2 Financial Impacts
The initial cost of the fuse is less than that of the MCB. However, the maintenance and replacement costs of the fuse eventually exceed the cost of the MCB. This therefore implies that the financial impact of the fuse is more severe than the one of the MCB.

#### 5.2.3 Short circuit withstands
The fuse can withstand only one short circuit, while the MCB can withstand three to five short circuits [2].

#### 5.2.4 Environmental Impacts
Emanuelson et al [3] concluded that porcelain (MCB) has no known emissions and it can be used as landfill.

**Fuse LCA Score:** 0
**MCB LCA Score:** 1

### 5.3 Speed, Quality and Cost (SQC) Analysis
The fuse reacts to a fault faster than the MCB because of its low current cut-off value. It can be concluded that the MCB has more quality than the fuse because of the many advantages it has over the fuse. The MCB has a longer lifespan than the fuse. This therefore implies that it is more cost-effective when compared to the fuse. This is due to the fact that the fuse has a shorter lifespan, and hence has to be regularly replaced, and the cost of replacement accumulates continually.

**Fuse SQC Score:** 0
**MCB SQC Score:** 1

## 6 Results and Discussion
The results from the individual analysis paradigms are converged are converged into the AND gate structure. According to table 1, the fuse is classified as bad and the MCB is classified as good. The terms "good" and "bad" are not used to classify these objects in a general sense, but an object is "bad" relative to its counterpart, and the other way round. The mathematics behind the classification process is shown below.

For the fuse, $y = 0\,AND\,0\,AND\,0 = 0$

For the MCB, $y = 1\,AND\,1\,AND\,1 = 1$

This implies that the MCB is the most reliable in terms of providing protection against overload currents and short circuit currents.

## 7 Conclusion
In this paper, two protection devices were analyzed for low voltage applications. These are the miniature circuit breaker (MCB) and the high rupturing capacity (HRC) fuse. A classifier approach was adopted for the analysis. A new tool called the 3-AND Convergence Classifier was introduced and tested. Making use of this tool, it was demonstrated and observed that the MCB is a "good" device for protection against the effects of over currents. This includes overload currents and short circuit currents.

It is also acknowledged that the results obtained from using the 3-AND Convergence Classifier may vary depending on what the analyst considers as an advantage or a disadvantage, according to their application.

One way of improving the efficiency of protection devices would be to built models of these devices and use artificial intelligence techniques to forecast and predict possible faults and hence be able to put measures in place to prevent them.

## 8 Acknowledgements
The author hereby thanks Rand Water's Electrical Engineering Section for the opportunity to do the work.